\documentclass[12pt]{iopart}
\usepackage{latexsym}

\def\tr{{\rm tr}}

\def\ket#1{\mid~\!\!\!{#1}~\!\!\rangle}
\def\bra#1{\langle~\!\!{#1}~\!\!\!\mid}
\def\cH{{\cal H}}

\def\IF{if and only if }

\def\qm{quantum mechanics}
\def\Q{quantum }

\def\QMl{quantum-mechanical }

\def\cR{{\cal R}}

\def\${\enskip$}
\def\M{measurement }
\def\m{measurement}

\begin{document}\jl{1}

\title[\bf Steering]{\bf
Schr\"{o}dinger's pure-state steering
completed}

\author{F Herbut\footnote[1]{E-mail:
fedorh@infosky.net and
fedorh@mi.sanu.ac.yu}}

\address{Serbian Academy of
Sciences and Arts, Knez Mihajlova 35,
11000 Belgrade, Serbia}

\date{\today}

\begin{abstract}\noindent Schr\"{o}dinger
investigated entanglement in
two-particle state vectors by assuming
measurement finding out if the nearby
particle is in a given state vector
\$\psi_1\$ or not. Without interaction
with the distant particle, just on
account of the entanglement, the
distant particle is steered into a
certain state vector. In
Schr\"{o}dinger's finite-dimensional
case thus any distant-particle state
vector can be reached. This theory was
extended to infinite-dimensional spaces
by the author. The present article
completes the extension by throwing
light on the fine structure of
steering.
\end{abstract}

PACS numbers: 03.65.-w, 03.65.Ca,
03.65.Db, 03.65.Ud

\maketitle \normalsize \rm

\section{Introduction}

When in 1935 Einstein et al. launched
their revolutionary EPR paradox
\cite{EPR} , many deep-thinking
foundationally-minded physicists
followed suit. Among them were Furry
\cite{Furry} and Schr\"{o}dinger
\cite{Schroed35}, \cite{Schroed36}. The
latter author introduced the now widely
used concept of entanglement, but also
that of disentanglement and of steering
or distant steering \cite{Spekkens}.
Schr\"{o}dinger's approach and
indignation can be seen in his words
\cite{Schroed35} p. 556: "It is rather
discomforting that the theory should
allow a system to be steered or piloted
into one or the other type of state at
the experimenter's mercy in spite of
his having no access to it." This is
made even more clear in his next paper
\cite{Schroed36}, p. 446: "... in
general a sophisticated experimenter
can, by a suitable device which does
not involve measuring non-commuting
variables, produce a non-vanishing
probability of driving the system into
any state he chooses". He had
two-particle pure states with
non-singular reduced density operators,
and finite dimensional state spaces of
particles in mind. (This will be
obvious after the detailed study in
this article.)

Distant steering in case of
two-particle state vectors that have
reduced density operators with
infinite-dimensional ranges were
studied by the present author
\cite{MVFH88}, \cite{FHJMP}. The
present
 article is actually a
completion of the former study with
insight in the fine structure of
steering. Wiseman et al. extended
steering to mixed two-particle states
\cite{Wiseman}.

This study is focused on {\it
two-particle state vectors} that have
{\it infinite-dimensional ranges} of
reduced density operators. (The theory
is general, but the fine structure
studied does not show up in the
trivial, finite-dimensional case.)\\

\section{The distant state}

It is well known that one of the basic
\QMl relations is the so-called {\it
trace rule}, which expresses the
probability \$p(P,\rho )\$ of
occurrence of a \Q event (projector)
\$P\$ in a \Q state (density operator)
\$\rho\$ by the simple formula
\$p(P,\rho )=\tr(P\rho )\$.
'Occurrence' is defined by \m , but it
is an astonishing \QMl fact (that one
is usually not aware of) that this
notion has a 'two-dimensional
multitude':

(i) One can take any observable
(Hermitian operator) \$A\$ of which
\$P\$ is an eigen-projector
corresponding to an (arbitrary)
eigenvalue \$a\$, i. e, an operator the
spectral form of which is
\$A=aP+P^{\perp}AP^{\perp}\$ where
\$P^{\perp}\equiv 1-P\$ (and the second
term does not have the eigenvalue
\$a\$). If in the \M of \$A\$ the
result \$a\$ is obtained, then one says
that \$P\$ has occurred.

(ii) The observable specified in (i)
can be measured in whatever way:
ideally (the textbook case), when the
L\"{u}ders formula \cite{Lud} gives the
change of state; in more general
non-demolition \M (older synonyms:
repeatable \m , or \M of the first
kind); in \M in which the result is not
preserved (non-repeatable or
second-kind \m ).

It is also not widely known that if one
has a bipartite system in any
correlated state (density operator)
\$\rho_{12}\$, i. e., when
\$\rho_{12}\not=\rho_1\otimes\rho_2\$,
where the tensor factors are the
reduced density operators, then, if a
first-subsystem event \$P_1\$ occurs
(in the sense defined in the preceding
passages with the two multitudes of
varieties), then the second subsystem
{\it ipso facto}, i. e., without any
interaction between the measuring
instrument and the subsystem, makes
transition from the subsystem state
(reduced density operator)
\$\rho_2\equiv\tr_1\rho_{12}\$ to the
following state (density operator) in
\$\cH_2\$:
%%%%%%%%%%%%%%%%%%%%%%%%%%%%%%%%%%%%%%
$$ p^{-1}\tr_1
\Big(\rho_{12}P_1 \Big),\eqno{(1a)}$$
where $$p\equiv\tr_{12}(P_1\rho_{12})
\eqno{(1b)}$$
%%%%%%%%%%%%%%%%%%%%%%%%%%%%%%%%%%%%%%
is the probability of the occurrence of
\$P_1\$ in the state \$\rho_{12}\$. We
write under the partial trace \$P_1\$
instead of \$P_1\otimes I_2\$, where
\$I_2\$ is the identity operator in
\$\cH_2$. (A proof of (1a,b) is given
in \cite{FH69}, subsection 6.B.)\\

We need two steps of confining
ourselves to special cases from
relation (1a) to come to
Schr\"{o}dinger's steering. We want to
do this in the antilinear
representation of bipartite state
vectors (vectors of norm one)
\cite{FHJMP}. (The indices \$1\$,
\$2\$, and \$12\$ show in which space
the entity is.)\\

\section{The antilinear representation}

There is an isomorphism from the tensor
product \$\cH_1\otimes\cH_2\$, where
the factors are complex separable
(finite or countably infinite
dimensional) Hilbert spaces (state
spaces of the subsystems) to {\it
antilinear Hilbert-Schmidt operators}
\$A_a\$ that map \$\cH_1\$ into
\$\cH_2\$, determined by {\it partial
scalar product}:
%%%%%%%%%%%%%%%%%%%%%%%%%%%%%%%%%%%%%%
$$\forall
\ket{\Phi}_{12}\quad\rightarrow\quad
A_a:\eqno{(2a)}$$ $$\forall
\ket{\psi}_1:\quad \Big(A_a\ket{\psi
}_1\Big)_2\equiv\bra{\psi
}_1\ket{\Phi}_{12},\eqno{(2b)}$$ where
\$\bra{\psi }_1\ket{\Phi}_{12}\$ is the
{\it partial scalar product} over
subsystem \$1$.
%%%%%%%%%%%%%%%%%%%%%%%%%%%%%%%%%%%%%%

Each antilinear operator \$A_a\$
defined by (2b) determines its adjoint
\$A_a^{\dag}\$, which maps \$\cH_2\$
into \$\cH_1\$. The adjoint is
determined via the relation
%%%%%%%%%%%%%%%%%%%%%%%%%%%%%%%%%%%%%%
$$\forall
\ket{\psi}_1,\enskip\ket{\phi}_2:\quad
\Big(A_a\ket{\psi}_1,\ket{\phi}_2\Big)_2=
\Big(\ket{\psi}_1,A_a^{\dag}\ket{\phi}_2
\Big)_1^*,\eqno{(3)}$$
%%%%%%%%%%%%%%%%%%%%%%%%%%%%%%%%%%%%%%
where the brackets stand for scalar
products, which are antilinear in the
first factor, and the asterisk denotes
complex conjugation.

Relation (2b) also implies \$\tr\Big(
A_a^{\dag}A_a\Big)<\infty\$. This is
the relation that makes the antilinear
operators \$A_a\$ and \$A_a^{\dag}\$
Hilbert-Schmidt ones.\\

\section{Two steps of special cases\\}

Now we make the first step of taking a
special case of (1a,b). By
\$P_{\chi_i}\$ we denote the projector
onto the one-dimensional subspace
spanned by the unit vector \$\chi_i\$,
and \$i=1,2,12\$ keeps track of the
Hilbert space to which the entity
belongs (even if it is superfluous, it
is useful for transparency). Naturally,
\$P_{\chi_i}\psi_i=\chi_i\Big(\chi_i,
\psi_i\Big)_i,\enskip i=1,2,12\$.\\

{\bf Theorem 1.} If one has any
bipartite {\it state vector}
\$\Phi_{12}\$, i. e., \$\rho_{12}\equiv
P_{\Phi_{12}}\$, and one goes over to
the antilinear representation \$A_a\$
of \$\Phi_{12}\$, the occurrence of any
first-subsystem event \$P_1\$ brings
about the following second-subsystem
state:
%%%%%%%%%%%%%%%%%%%%%%%%%%%%%%%%%%%%%%
$$p^{-1}\Big[A_aP_1A_a^{\dag}\Big]_2,
\eqno{(4a)}$$ where $$p\equiv\Big(
\Phi_{12},P_1\Phi_{12}\Big)_{12}
\eqno{(4b)}$$
%%%%%%%%%%%%%%%%%%%%%%%%%%%%%%%%%%%%%%
is the probability of the event \$P_1\$
in the state \$\Phi_{12}$.\\

{\bf Proof.} As to the antilinear
representation, we are going to utilize
(2b) and (3), and the fact that numbers
undergo complex conjugation when taken
to the left from an antilinear
operator. Let \$\{\phi_2^n:\forall
n\}\$ be a complete orthonormal basis
in \$\cH_2\$, and let \$\{\psi_1^k:
\forall k\}\$ be a complete orthonormal
basis in \$\overline{\cR(P_1)}\$, the
topological closure of the range of
\$P_1\$: \$P_1=\sum_kP_{\psi_1^k}\$.
Then
$$\Big(\phi_2^n,
\Big[\tr_1(P_{\Phi_{12}}P_1)\Big]_2
\phi_2^{n'}\Big)_2=\sum_k\Big(\psi_1^k
\phi_2^n,P_{\Phi_{12}}\psi_1^k\phi_2^{n'}
\Big)_{12}=$$
$$\sum_k\Big(\psi_1^k\phi_2^n,\Phi_{12}
(\Phi_{12},\psi_1^k\phi_2^{n'})_{12}\Big)
_{12}=\sum_k\Big(\psi_1^k\phi_2^{n'},
\Phi_{12}\Big)_{12}^*\Big(\phi_2^n,A_a
\psi_1^k\Big)_2=$$
$$\sum_k\Big(\phi_2^{n'},A_a\psi_1^k\Big)
_2^*\Big(\phi_2^n,A_a\psi_1^k\Big)_2=
\sum_k\Big(A_a^{\dag}\phi_2^{n'},
\psi_1^k\Big)_1
\Big(\phi_2^n,A_a\psi_1^k\Big)_2=$$
$$\sum_k\Big(\psi_1^k,
A_a^{\dag}\phi_2^{n'}\Big)_1^*
\Big(\phi_2^n,A_a\psi_1^k\Big)_2=
\Big(\phi_2^n,\bigg[A_a\sum_kP_{\psi_1^k}
\Big[A_a^{\dag}\phi_2^{n'}\Big]_1
\bigg]_2\Big)_2=$$
$$\Big(\phi_2^n,\bigg[A_aP_1
\Big[A_a^{\dag}\phi_2^{n'}
\Big]_1\bigg]_2\Big)_2=
\Big(\phi_2^n,\Big[A_aP_1A_a^{\dag}
\Big]_2\phi_2^{n'}\Big)_2.$$\hfill
$\Box$\\

The second step of taking a special
case of (4a,b) is confining ourselves
to {\it elementary} first-subsystem
{\it events} (ray projectors)
\$P_1\equiv P_{\psi_1}\$, where
\$\psi_1\$ is an arbitrary state
vector.\\

{\bf Theorem 2.} If an elementary event
\$P_{\psi_1}\$ occurs on the first
subsystem in a state vector
\$\Phi_{12}\$ (or \$A_a\$), then the
second subsystem finds itself in the
state described by the {\it state
vector}
%%%%%%%%%%%%%%%%%%%%%%%%%%%%%%%%%%%%%%
$$A_a\psi_1\Big/
||A_a\psi_1||,\eqno{(5a)}$$ and the
{\it probability} of the occurrence of
\$P_{\psi_1}\$ is $$p=||A_a\psi_1||^2.
\eqno{(5b)}$$
%%%%%%%%%%%%%%%%%%%%%%%%%%%%%%%%%%%%%%

{\bf Proof.} As to the claimed
probability (5b), from (4b) one has
$$p=\tr_{12}\Big(
P_{\Phi_{12}}P_{\psi_1}\Big).$$
Introducing complete orthonormal bases
\$\{\psi_1^k:\forall
k;\psi_1^{k=1}\equiv\psi_1\}\$ and
\$\{\phi_2^n:\forall n\}\$, one obtains
$$p=\sum_n\Big(\psi_1\phi_2^n,
P_{\Phi_{12}}
\psi_1\phi_2^n\Big)_{12}.$$ Applying
the projector, and taking out one
scalar product from the other, one
further has
$$p=\sum_n
\Big(\Phi_{12},\psi_1\phi_2^n\Big)_{12}
\Big(\psi_1\phi_2^n,\Phi_{12}\Big)_{12}=$$
$$\sum_n\Big(\psi_1\phi_2^n,\Phi_{12}\Big)
_{12}^*\Big(\phi_2^n,A_a\psi_1\Big)_2=
\sum_n\Big(\phi_2^n,A_a\psi_1\Big)_2^*
\Big(\phi_2^n,A_a\psi_1\Big)_2=||A_a\psi_1
||^2.$$

To derive claim (5a), we start with
(4a), and we utilize the above basis in
\$\cH_2\$.
$$\Big(\phi_2^n,p^{-1}\Big[A_aP_{\psi_1}
A_a^{\dag}\Big]_2\phi_2^{n'}\Big)_2=
\Big(\phi_2^n,\Big[p^{-1/2}A_a\psi_1\Big]_2
(\psi_1,[p^{-1/2}A_a^{\dag}
\phi_2^{n'}]_1)_1\Big)_2 =$$
$$\Big(\psi_1,\Big[p^{-1/2}A_a^{\dag}
\phi_2^{n'}\Big]_1\Big)_1^*\Big(\phi_2^n,
\Big[p^{-1/2}A_a\psi_1\Big]_2\Big)_2=
\Big(\Big[p^{-1/2}A_a\psi_1\Big]_2,
\phi_2^{n'}\Big)_2\Big(\phi_2^n,
\Big[p^{-1/2}A_a\psi_1\Big]_2\Big)_2=$$
$$\Big(\phi_2^n,\Big[p^{-1/2}A_a\psi_1
\Big]_2 ([p^{-1/2}A_a\psi_1]_2,
\phi_2^{n'})_2\Big)=
\Big(\phi_2^n,P_{[p^{-1/2}A_a\psi_1]_2}
\phi_2^{n'}\Big)_2.\qquad\Box$$ One
should note that after the second
equality the scalar product is complex
conjugated because before the first
equality, \$A_a\$ is seen to act after
the projector, hence also on the
numbers that come out as a result of
the projection. Contrariwise, if we
read the next to last expression one
step backwards, the scalar product is
extracted without complex conjugation
though it is to the right of \$A_a\$.
The reason is that \$A_a\$ acts in
\$\cH_1\$, and the scalar product (a
number) appears in \$\cH_2\$ after the
action of \$A_a\$.\\

{\it Schr\"{o}dinger's steering} is
defined for the occurrence of an
arbitrary elementary first-subsystem
event \$P_{\psi_1}\$ in an arbitrary
bipartite state vector \$\Phi_{12}\$.
As it is shown in Theorem 2, this boils
down to {\it mapping \$\cH_1\$ into
\$\cH_2$ by \$A_a\$} (the antilinear
representative of \$\Phi_{12}\$).\\

\section{Polar factorization\\}

To bring out the full power of the
antilinear representation, one should
perform the two polar factorizations of
\$A_a\$ \cite{FHMV76}:
%%%%%%%%%%%%%%%%%%%%%%%%%%%%%%%%%%%%%%
$$ A_a=U_a\rho_1^{1/2},
\eqno{(6a)}$$
$$A_a=\rho_2^{1/2}U_aQ_1,\eqno{(6b)}$$
%%%%%%%%%%%%%%%%%%%%%%%%%%%%%%%%%%%%%%
where \$\rho_i\equiv
\tr_jP_{\Phi_{12}}\$, \$i,j=1,2,\enskip
i\not= j\$ is the \$i$-th subsystem
state (reduced density operator), and
\$U_a\$ is an antilinear unitary
operator mapping the topologically
closed range \$\overline{\cR(\rho_1)}\$
onto the topologically closed range
\$\overline{\cR(\rho_2)}\$ (these
subspaces are always equally
dimensional), and, finally, \$Q_1\$ is
the range-projector of \$\rho_1$.

The operator \$U_a\$ is called {\it the
correlation operator}. It is the only
precise mathematical entity expressing
the quantum correlations inherent in a
bipartite state (known to the author).\\

{\bf Remark 1.} As it is seen in (6b),
Schr\"{o}dinger's steering maps
\$\cH_1\$ into \$\cR(\rho_2^{1/2})\$.
Actually, it is a surjection, i. e., an
"onto" map \cite{MVFH88}. This is, of
course, non-trivial only in case of
infinite-dimensional ranges (of
\$\rho_i,\enskip i=1,2\$), when one
should have in mind the known proper
inclusion relations: $$\cR(\rho
)\subset\cR(\rho^{1/2})\subset
\overline{\cR(\rho )}\eqno{(7)}$$ valid
for any density operator with
infinite-dimensional range.\\

\section{Largest probability of
steering\\}

We proceed by analyzing (5a,b) to gain
detailed insight in Schr\"{o}dinger's
steering.

{\bf Theorem 3. A)} Two first-subsystem
state vectors \$\psi_1\$ and
\$\psi_1'\$ give, upon \m , the {\it
same steering} in subsystem \$2\$ \IF

(i) the range-projections are
positively collinear: $$Q_1\psi_1=
cQ_1\psi_1',\quad c>0,\eqno{(8a)}$$ or
{\it equivalently}

(ii) if they determine by projection
the same state vector in \$\overline{
\cR(\rho_1)}\$: $$Q_1\psi_1\Big/
||Q_1\psi_1||=Q_1\psi_1'\Big/
||Q_1\psi_1'||.\eqno{(8b)}$$

{\bf B)} Of all elementary events
\$P_{\psi_1}\$ in \$\cH_1\$ that give
one and the same state vector in
\$\cH_2\$ by steering {\it largest
probability} of occurrence has the one
that lies entirely in \$\overline{
\cR(\rho_1)}\$, or, equivalently, the
component of which in the null
space of \$A_a\$ is zero.\\

{\bf Proof. A)} {\it Sufficiency.} Let
(8a) be valid. Since
\$\rho_1=\rho_1Q_1\$, and
\$\rho_1^{1/2}=\rho_1^{1/2}Q_1\$, (6a)
implies \$A_a=A_aQ_1\$. Hence, applying
\$A_a\$ to (8a), one obtains
\$A_a\psi_1=cA_a\psi_1'\$, and
\$||A_a\psi_1||=c||A_a\psi_1'||\$.
Finally, $$A_a\psi_1\Big/||A_a\psi_1||=
A_a\psi_1'\Big/||A_a\psi_1'||.\eqno{(9)}$$

{\it Necessity.} If relation (9) is
valid, then $$A_a\Big(\psi_1\Big/
||A_a\psi_1||-\psi_1'\Big/
||A_a\psi_1'||\Big)=0=Q_1
\Big(\psi_1\Big/
||A_a\psi_1||-\psi_1'\Big/
||A_a\psi_1'||\Big)$$ (\$A_a\$ and
\$Q_1\$ have the same null space).
Finally, $$Q_1\psi_1=\Big(||A_a\psi_1||
\Big/||A_a\psi_1'||\Big)Q_1\psi_1'.$$
Thus, (8a) is satisfied.

Clearly, (8b) implies (8a). Conversely,
(8a) gives \$||Q_1\psi_1||=c
||Q_1\psi_1'||\$. Relation (8b) ensues
from (8a) and this relation.

{\bf B)} Relation (5b) implies
$$p=||A_a\psi_1||^2=||A_aQ_1\psi_1||^2=
\Big(||Q_1\psi_1||^2\Big)\Big[
||A_a\Big(
Q_1\psi_1\Big/||Q_1\psi_1||\Big)||^2\Big].
\eqno{(10)}$$ All vectors specified in
(8b) have the second factor after the
last equality in (10) in common.
Therefore, the probability is largest
when the first factor (after the last
equality in (10)) is largest, i. e.,
when it is one.\hfill $\Box$\\

\section{The fine structure of
infinite-dimensional ranges}

In this section we make a deviation
from our two-particle study to one
Hilbert space and a given density
operator with an infinite-dimensional
range in it.

{\bf Remark 2.} Let \$\rho\$ be a
density operator with an
infinite-dimensional range. Writing
"\$\oplus\$" for the union of disjoint
sets, and "\$ \ominus\$" when
set-theoretically subtracting a subset
from a larger set, the proper-inclusion
chain (7) implies
$$\overline{\cR(\rho )}=\cR(\rho )
\oplus \Big(\cR(\rho^{1/2})\ominus
\cR(\rho )\Big)\oplus\Big(
\overline{\cR(\rho
)}\ominus\cR(\rho^{1/2})\Big).
\eqno{(11)}$$\\

{\bf Lemma 1.} Let \$\{\psi_k:\forall
k\}\$ be a complete orthonormal
eigenbasis of \$\rho\$ in
\$\overline{\cR(\rho )}\$, and let
\$\{r_k:\forall k\}\$ be the
corresponding positive spectrum of
\$\rho\$ (with possible repetitions of
equal eigenvalues in general). Let,
further, \$\overline{\cR(\rho
)}\enskip\ni \enskip\psi
=\sum_ka_k\psi_k\$, with \$\forall
k:\enskip a_k\in\mbox{\bf C}\$ be an
arbitrary element, i. e.,
\$\sum_k|a_k|^2<\infty\$. Then $$\psi
\in\cR(\rho )\quad\Leftrightarrow \quad
\sum_k|r_k^{-1}a_k|^2<\infty,
\eqno{(12)}$$ and $$\psi
\in\cR(\rho^{1/2})\quad\Leftrightarrow
\quad \sum_k|r_k^{-1/2}a_k|^2<\infty.
\eqno{(13)}$$

{\bf Proof.} {\it \$\{\Leftarrow\$ in
(12)\}.} Assuming the validity of the
second expression in (12), we define
\$\phi\equiv\sum_k r_k^{-1}a_k\psi_k\$.
Then one has \$\psi = \rho\phi\$, i.
e., the first expression in (12) holds
true.

{\it \$\{\Rightarrow\$ in (12)\}.} If
\$\psi\$ belongs to the range, there
exists \$\phi =\sum_kb_k\psi_k,\enskip
\sum_k|b_k|^2<\infty\$, and \$\rho\phi
=\psi\$. Since \$\forall k:\enskip
a_k=r_kb_k\$, one has \$\forall
k:\enskip\sum_k|r_k^{-1}a_k|^2<\infty$.

Equivalence (13) is proved analogously.
\hfill $\Box$\\

{\bf Lemma 2.} The square root
\$\rho^{1/2}\$ of any density operator
\$\rho\$ with an infinite-dimensional
range maps in an one-to-one way
\$\overline{\cR(\rho )}\$ onto
\$\cR(\rho^{1/2})\$, and by this it
maps \$\cR(\rho^{1/2})\$ onto
\$\cR(\rho )\$, and
\$\Big\{\overline{\cR(\rho
)}\ominus\cR(\rho^{1/2})\Big\}\$ onto
\$\Big\{\cR(\rho^{1/2})\ominus \cR(\rho
)\Big\}\$, i. e., (14a-c) is valid:

$$\quad\quad\quad\cR\Big(\rho^{1/2}\Big)
\oplus\bigg\{\overline{\cR(\rho )}
\ominus\cR\Big(\rho^{1/2}\Big)\bigg\}
=\overline{\cR(\rho )}\eqno{(14a)}$$
$$\downarrow\quad\quad\quad\quad\quad\quad
\quad\downarrow\quad\quad\quad
\eqno{(14b)}$$
$$\quad\quad\quad\quad\cR(\rho )
\enskip\oplus
\enskip\bigg\{\cR(\rho^{1/2})
\ominus\cR(\rho )\bigg\}=
\cR(\rho^{1/2}).\eqno{(14c)}$$\\

{\bf Proof.} That \$\rho^{1/2}\$ maps
\$\overline{\cR(\rho )}\$ into
\$\cR(\rho^{1/2})\$ is obvious from
(6b). To prove that it is an "onto"
map, let \$\psi=\sum_ka_k\psi_1\$ (cf
Lemma 1) be an arbitrary element of the
latter range. Then, according to (13),
also \$\phi\equiv\sum_kr_k^{-1/2}a_k
\psi_k\$ is an element of
\$\overline{\cR(\rho )}\$. Applying
\$\rho^{1/2}\$ to it, we obtain
\$\psi\$. Assuming {\it ab contrario}
that \$\phi ,\phi'\in
\overline{\cR(\rho )}\$, \$\phi\not=
\phi'\$, and \$\rho^{1/2}\phi =
\rho^{1/2}\phi'\$, one arrives at
\$\rho^{1/2}(\phi -\phi')=0\$, i.e., a
non-zero element is taken into zero.
This is not possible because
\$\rho^{1/2}\$ has the same null space
as \$\rho\$, and it is the
orthocomplement of \$\overline{\cR(\rho
)}$.

The first arrow in (14b), i. e., the
map \$\rho^{1/2}\$ that it denotes, is
obvious in the "into" sense because
\$\rho^{1/2}\rho^{1/2}=\rho\$. Let
\$\psi=\sum_ka_k\psi_k\$ be an
arbitrary element of \$\cR(\rho )\$.
Then, according to (12),
\$\sum_k|r_k^{-1}a_k|^2<\infty\$. Then
also \$\sum_k|r_k^{-1/2}a_k|^2<\infty\$
(compare the first inclusion in (7)
with (12) and (13)). Hence, we can
define \$\phi\equiv
\sum_kr_k^{-1/2}a_k\$, and we have
\$\rho^{1/2}\phi =\psi\$. Thus, we are
dealing with an "onto" map.

Finally, the last claim is an immediate
consequence of the preceding two, as
easily seen.\hfill $\Box$\\

\section{Back to steering}

It was shown in previous work
\cite{FHMV76} that the correlation
operator not just maps
\$\overline{\cR(\rho_1)}\$ onto
\$\overline{\cR(\rho_2)}\$. It takes by
similarity transformation the positive
part of one reduced density operator
into that of the other: $$\rho_2=
U_a\rho_1U_a^{-1}Q_2,\eqno{(15)}$$
where \$Q_2\$ is the range projector of
\$\rho_2$.

{\bf Lemma 3.} The correlation operator
preserves decomposition (11):
$$U_a\overline{\cR(\rho_1)}=
\overline{\cR(\rho_2)},\eqno{(16a)}$$
$$U_a\cR(\rho_1)=\cR(\rho_2),\eqno
{(16b)}$$
$$U_a\Big(\cR(\rho_1^{1/2})\ominus
\cR(\rho_1)\Big)=
\Big(\cR(\rho_2^{1/2})\ominus
\cR(\rho_2)\Big),\eqno{(16c)}$$
$$U_a\Big(\overline{\cR(\rho_1)}
\ominus\cR(\rho_1^{1/2})\Big)=\Big(
\overline{\cR(\rho_2 )}
\ominus\cR(\rho_2^{1/2})\Big).\eqno{(
16d)}$$\\

{\bf Proof.} In Lemma 1 we made the
choice \$\forall k:\enskip \rho_1
\psi_1^k=r_k\psi_1^k\$. Applying the
correlation operator, one obtains
\$\forall k:\enskip
\Big(U_a\rho_1U_a^{-1}\Big)
\Big(U_a\psi_1^k\Big)=r_k\Big(U_a
\psi_1^k\Big)\$. Defining \$\forall
k:\enskip \psi_2^k\equiv U_a\psi_1^k\$,
one can, account of (15), write
\$\forall k:\enskip
\rho_2\psi_2^k=r_k\psi_2^k\$. Since
\$\cR(\rho_i) \enskip i=1,2\$ is the
linear manifold spanned by the
eigenvectors \$\{\psi_i^k:\forall k\},
\enskip i=1,2\$, (16b) is valid.

According to (13), \$\sum_k|r_k^{-1/2}
a_k|^2<\infty\$ is satisfied for every
element \$\psi_1=\sum_ka_k\psi_1^k\$
that belongs to \$\cR(\rho_1^{1/2})\$.
Applying \$U_a\$, one has \$\Big(U_a
\psi_1\Big)=\sum_ka_k^*\psi_2^k\$, and
\$\sum_k|r_k^{-1/2}a_k^*|^2=
\sum_k|r_k^{-1/2} a_k|^2<\infty\$.
Thus, \$U_a\cR(\rho_1^{1/2})=
\cR(\rho_2^{1/2})\$. The rest in the
claim
 is evident.\hfill $\Box$\\

{\bf Theorem 4.} The antilinear
representative \$A_a\$ of a given
bipartite state vector \$\Phi_{12}\$
(cf (2a,b)) that implies reduced
density operators with
infinite-dimensional ranges maps in an
one-to-one way \$\overline{\cR(\rho_1
)}\$ onto \$\cR(\rho_2^{1/2})\$, and by
this it maps \$\cR(\rho_1^{1/2})\$ onto
\$\cR(\rho_2)\$, and
\$\Big\{\overline{\cR(\rho_1
)}\ominus\cR(\rho_1^{1/2})\Big\}\$ onto
\$\Big\{\cR(\rho_2^{1/2})\ominus
\cR(\rho_2)\Big\}\$. This is made more
transparent by the following relations:
$$\quad\quad\quad\cR\Big(\rho_1^{1/2}\Big)
\oplus\bigg\{\overline{\cR(\rho_1)}
\ominus\cR\Big(\rho_1^{1/2}\Big)\bigg\}
=\overline{\cR(\rho_1)}\eqno{(17a)}$$
$$\downarrow\quad\quad\quad\quad\quad\quad
\quad\downarrow\quad\quad\quad
\eqno{(17b)}$$
$$\quad\quad\quad\quad\cR(\rho_2)
\enskip\oplus
\enskip\bigg\{\cR(\rho_2^{1/2})
\ominus\cR(\rho_2)\bigg\}=
\cR(\rho_2^{1/2}).\eqno{(17c)}$$\\

{\bf Proof.} The claim of the theorem
is evident having in mind the polar
factorization (6b) of \$A_a\$, Lemma 3
and Lemma 2 in application to
\$\rho_2$.\hfill $\Box$\\

\section{Conclusion} In the formalism
t
he map \$A_a\$ that represents
antilinearly any given bipartite state
vector \$\Phi_{12}\$ {\it performs
actually the Schr\"{o}dinger steering}.
If the composite-system state vector
implies infinite-dimensional reduced
density operators \$\rho_i,\enskip
i=1,2\$, then the mapping has a fine
structure:

(i) If the \M of \$\psi_1\$ is that of
an eigenvector of \$\rho_1\$
corresponding to a positive eigenvalue,
then actually the corresponding
eigenvector \$\psi_2=U_a\psi_1\$ is
{\it distantly measured}. This simplest
case was extensively studied in
\cite{FHMV76} and \cite{MVFH84} in the
non-selective version of \m , when all
results are taken into account in
contrast to Schr\"{o}dinger's steering,
in which the selective version of \M is
considered with only one result - that
of obtaining \$1\$ for \$P_{\psi_1}$.

(ii) All other elements of
\$\cR(\rho_2)\$ can be obtained by
steering that results from direct \M of
a vector \$\psi_1\$ from
\$\cR(\rho_1^{1/2})\$. This case was
studied in detail, again in the
non-selective version of \m , in
\cite{FHJMP}. Finally:

(iii) The elements of
\$\Big(\cR(\rho_2^{1/2})\ominus
\cR(\rho_2)\Big)\$ can be reached by
steering when direct \M of vectors from
\$\Big(\overline{\cR(\rho_1)}\ominus
\cR(\rho_1^{1/2})\Big)\$ is performed.

Besides, the vectors from the range of
\$\rho_1\$ give, by selective \m , the
largest probability. Hence, the null
space of \$\rho_1\$ is best discarded
in steering.

The paradoxical physical meaning of
distant steering is not discussed in
this article. Quantum-mechanical
insight in the nice EPR-type
entanglement experiments of Scully et
al. \cite{Scully1}, \cite{Scully2} (a
thought and a real experiment) gained
by the present author recently
\cite{FHScully1}, \cite{FHScully2} has
led to the conclusion that distant
correlations are paradoxical only in
the Einsteinian absolute-property
interpretation of \qm . If one takes
resort to the alternative, the
relative-property interpretation, a
kind of Everettian approach, then
nothing is paradoxical.\\

\end{document}